\documentclass[%
prl,
reprint,
amsmath,
amssymb,
aps,
]{revtex4-2}

\usepackage{graphicx}% Include figure files
\usepackage{bm}% bold math
\usepackage{physics}
\usepackage{tikz}
\usepackage{pgfplots}
\usepackage{hyperref}
\usepackage{url}
\usetikzlibrary{decorations.pathmorphing, positioning, shapes,arrows,arrows.meta}
\usetikzlibrary{shapes.geometric}
\usetikzlibrary{fadings}
\usepgfplotslibrary{fillbetween}
\tikzfading %strangely gives bad bounding box when inside the tikzpicture
[
  name=fade out,
  inner color=transparent!0,
  outer color=transparent!100
]
\usepackage[final]{changes}
\usepackage{natbib}

\begin{document}
\title{Optimal observables and estimators for practical superresolution imaging}
\author{Giacomo Sorelli} 
\affiliation{Laboratoire Kastler Brossel, Sorbonne Universit\'e, ENS-Universit\'e  PSL, CNRS, Collège de France, 4  Place Jussieu, F-75252 Paris, France}
\author{Manuel Gessner}
%\affiliation{Laboratoire Kastler Brossel, Sorbonne Universit\'e, ENS-Universit\'e PSL, CNRS,Collège de France, 24 Rue Lhomond, 75005, Paris, France}
\affiliation{Laboratoire Kastler Brossel, Sorbonne Universit\'e, ENS-Universit\'e PSL, CNRS, Collège de France, 4  Place Jussieu, F-75252 Paris, France}
\author{Mattia Walschaers} 
\affiliation{Laboratoire Kastler Brossel, Sorbonne Universit\'e, ENS-Universit\'e PSL, CNRS, Collège de France, 4  Place Jussieu, F-75252 Paris, France}
\author{Nicolas Treps}
\affiliation{Laboratoire Kastler Brossel, Sorbonne Universit\'e, ENS-Universit\'e PSL, CNRS, Collège de France, 4  Place Jussieu, F-75252 Paris, France}

\date{\today}
\begin{abstract}
Recent works identified resolution limits for the distance between incoherent point sources. 
However, it remains unclear how to choose suitable observables and estimators to reach these limits in practical situations. 
Here, we show how estimators saturating the Cram\'er-Rao bound for the distance between two thermal point sources can be constructed using an optimally designed observable in the presence of practical imperfections, such as misalignment, crosstalk and detector noise.
\end{abstract}

\maketitle

{\it Introduction\:--}
Our capability of resolving small details in microscopy, and remote objects in astronomy is determined by the achievable precision of optical imaging.
Historical resolution limits -- as those of Abbe \cite{Abbe} or Rayleigh \cite{Rayleigh} -- address the effect of diffraction, but they can be overcome when the signal-to-noise ratio is high enough \cite{goodman2015statistical}.
In fact, the last decade provided us with a variety of superresolution techniques to beat these limits by fluorescence microscopy \cite{Hell:94, Klar8206, Betzig1642}, homodyne measurements \cite{Hsu_2004, Delaubert:06, PinelPRA2012} or intensity measurements in an appropriate basis \cite{Helstrom73,Tsang2019}. 

A paradigmatic imaging problem is the estimation of the separation of two incoherent point sources. 
By using tools from quantum metrology 
\cite{helstrom1976, BraunsteinCaves1994, holevo2011probabilistic,Paris2009,GiovannettiLoydMaccone,Luca_Augusto_review}, this can be optimally solved by spatial-mode demultiplexing \cite{Tsang_PRX}.
Extensions of these results to thermal sources \cite{Nair_2016, LupoPirandola}, to two-dimensional imaging \cite{Ang2017} and (for faint sources) to more general scenarios \cite{Rehacheck2017,Backlund2018,Yu2018, Napoli2019} are also available.

Several experiments \cite{Paur:16, Tang:16, Yang:16, Tham:2017} implemented a binary version of this demultiplexing technique, that distinguishes between the fundamental and the first excited modes. 
Modern light-shaping techniques, such as multi-plane-light-conversion \cite{Morizur:10} or wave-front shaping \cite{RotterGigan}, have recently enabled experiments demultiplexing into multiple modes \cite{Boucher:20}.
\replaced{In all these experiments, only the information obtained from a single mode (albeit up to 9 different ones in \cite{Boucher:20}) was used to estimate the parameter.}{However, in all experiments performed so far, the distance estimation was performed by looking at the intensity of one mode at the time.}
To push these experiments towards their ultimate resolution limits, it is crucial to determine a practical estimation strategy that optimally combines the information contained into all demultiplexed modes.

In this letter, we identify such an estimation strategy.
In particular, we show how an estimator for the separation between two, arbitrarily bright, thermal sources can be constructed using only the average of an optimized linear combination of demultiplexed intensity measurements.
Accordingly, this estimator is remarkably simpler to implement experimentally than standard methods requiring the full counting statistics.
Moreover, it takes into account misalignment  \cite{Tsang_PRX,Grace:20,AlmeaidaPRA2021}, crosstalk \cite{gessner2020} and detector noise \cite{Len2020,Lupo2020,oh2020}, and therefore it is directly relevant for practical applications.
Even in presence of the aforementioned imperfections, for faint sources, this estimator is efficient, i.e. it saturates the Cram\'er-Rao bound. 
Moreover, we demonstrate that our strategy is, in the noiseless case, also optimal for arbitrarily bright sources, if sufficiently many modes are measured.

{\it Source model\:--}
We estimate the transverse separation $d$ between two point sources located at positions $\pm{\bf r}_0$, with ${\bf r}_0 = (d\cos \theta, d \sin \theta)/2$.
After propagation through a diffraction-limited imaging system (possibly with finite transmissivity $\kappa$) the sources are described by the modes $u_0({\bf r} \pm {\bf r}_0)$, with $u_0({\bf r})$ the real point spread function (PSF) of the imaging system.
The modes $u_0({\bf r} \pm {\bf r}_0)$ are non orthogonal, therefore they cannot be used to properly represent the quantum state of the sources in the image plane. 
This issue is solved by introducing the orthogonal modes $u_\pm({\bf r}) = (u_0({\bf r} + {\bf r}_0) \pm u_0({\bf r} - {\bf r}_0))/\sqrt{2(1\pm \delta)}$, with 
\begin{equation}
\delta = \int d^2{\bf r} u^*_0({\bf r} + {\bf r}_0)u_0({\bf r} - {\bf r}_0),
\label{delta}
\end{equation}
the overlap between the images of the two sources.
The field operators $\hat{b}_{\pm}$ associated with the modes $u_\pm({\bf r})$ can be related to the operators $\hat{s}_{\pm}=(\hat{s}_1 \pm \hat{s}_2)/\sqrt{2}$, with $\hat{s}_{1/2}$ the field operators associated with the modes generated by the sources, according to \cite{LupoPirandola}
\begin{equation}
\hat{b}_{\pm} = \sqrt{\kappa_\pm}\hat{s}_\pm + \sqrt{1 -\kappa_{\pm}}\hat{v}_\pm,
\label{mapping}
\end{equation}
where $\hat{v}_\pm$ are field operators associated with auxiliary modes, that we can assume to be in the vacuum state, and $\kappa_{\pm} = \kappa(1 \pm \delta)$.
Equation \eqref{mapping} allows to propagate the quantum state $\hat{\rho}_0$ of the sources into its image $\hat{\rho}(d,\theta)$ after transmission through the imaging system \cite{sorelli2021moment}.

We assume the sources to be in the state $\hat{\rho}_0 = \hat{\rho}_{s_1}(N) \otimes \hat{\rho}_{s_2}(N)$, with $\hat{\rho}_{a}(N)$ a thermal state with mean photon number $N$ in the mode associated with the field operator $\hat{a}$. 
Thermal sources with unequal brightnesses are discussed in \cite{sorelli2021moment}.
According to Eq.~\eqref{mapping}, $\hat{\rho}_0$ is mapped into $ \hat{\rho}(\theta, d) = \hat{\rho}_{b_+}(N_+) \otimes \hat{\rho}_{b_-}(N_-)$ in the image plane, with the information on the parameter $d$ encoded both in the shape of the modes $u_\pm({\bf r})$ and the mean photon numbers $N_{\pm} = N \kappa_{\pm}$ \cite{sorelli2021moment}.

\begin{figure}[t]
\centering
%\documentclass{standalone}
%\usepackage{tikz}
%\usepackage{pgfplots}
%\usetikzlibrary{decorations.pathmorphing, positioning, shapes,arrows,arrows.meta}
%\usetikzlibrary{shapes.geometric}
%\usetikzlibrary{fadings}
%\usepgfplotslibrary{fillbetween}
%\tikzfading %strangely gives bad bounding box when inside the tikzpicture
%[
%  name=fade out,
%  inner color=transparent!0,
%  outer color=transparent!100
%]
%
%\begin{document}
\begin{tikzpicture}
\pgfmathdeclarefunction{gauss}{2}{%
  \pgfmathparse{1/(#2*sqrt(2*pi))*exp(-((x-#1)^2)/(2*#2^2))}%
}

% x axis
\draw[-{Latex},line width= 1 pt] (-3.8,1.2) -- (-3.8,3.2) node[right]{$d$};

% Gaussians
\begin{axis}[
	at ={(-3.8 cm, 2.5 cm)},
  	mark=none,
  	domain=-4:4,
  	samples=50,
  	smooth,
  	width=2 cm,
  	height=2 cm,
  	clip=false,
	axis y line=none,
	axis x line=none,
	ymin=0,
	xtick=\empty,
]
  \addplot [thick,red!70!black] ({gauss(-1,1)},{x});
  \addplot [thick,red!70!black] ({gauss(1,1)},{x});
\end{axis}

% misalignment
\draw[line width= 1 pt] (-3.8,2) -- (-2.8,2);
\draw[dashed] (-3.8,2.5) -- (-3.2,2.5);  
\draw[<->] (-3.3,2.5) -- node[right]{$d_s$}(-3.3,2.);

\begin{scope}[xshift = -0.3 cm]

%% crosstalk
\draw[fill=black] (-2.4,1.8) rectangle (-2.2,2.2);
\draw[line width= 1 pt] (-2.3, 0.5) rectangle (0, 3.5);
\draw (-2.1,3.) rectangle (-1.5,3.4); 
\node at (-1.8,3.2) {$u_{00}$};
\draw (-2.1,2.5) rectangle (-1.5,2.9); 
\node at (-1.8,2.7) {$u_{01}$};
\draw (-2.1,2.) rectangle (-1.5,2.4); 
\node at (-1.8,2.2) {$u_{10}$};
\node at (-1.8,1.6) {$\vdots$};
\draw (-2.1,0.6) rectangle (-1.5,1.); 
\node at (-1.8,0.8) {$u_{QQ}$};
\draw (-0.7,3.) rectangle (-0.1,3.4); 
\node at (-0.4,3.2) {$v_{00}$};
\draw (-0.7,2.5) rectangle (-0.1,2.9); 
\node at (-0.4,2.7) {$v_{01}$};
\draw (-0.7,2.) rectangle (-0.1,2.4); 
\node at (-0.4,2.2) {$v_{10}$};
\node at (-0.4,1.6) {$\vdots$};
\draw (-0.7,0.6) rectangle (-0.1,1.); 
\node at (-0.4,0.8) {$v_{QQ}$};
\draw[color = red!70!white, line width = 0.5 pt] (-1.5,3.2) -- (-0.7,2.7);
\draw[color = red!70!white, line width = 0.5 pt] (-1.5,2.2) -- (-0.7,2.7);
\draw[color = red!70!white, line width = 0.5 pt] (-1.5,2.2) -- (-0.7,3.2);
\draw[color = red!70!white, line width = 0.5 pt] (-1.5,3.2) -- (-0.7,2.2);
\draw[color = red!70!white, line width = 0.5 pt] (-1.5,2.7) -- (-0.7,3.2);
\draw[color = red!70!white, line width = 0.5 pt] (-1.5,2.7) -- (-0.7,2.2);
\draw[color=green!70!black, line width = 2 pt] (-1.5,3.2) -- (-0.7,3.2);
\draw[color=green!70!black, line width = 2 pt] (-1.5,2.7) -- (-0.7,2.7);
\draw[color=green!70!black, line width = 2 pt] (-1.5,2.2) -- (-0.7,2.2);
\draw[color=green!70!black, line width = 2 pt] (-1.5,0.8) -- (-0.7,0.8);

%% dark counts
\draw[line width = 1 pt] (-0.1, 3.2) -- (0.2, 3.2);
\draw[line width = 1 pt] (-0.1, 2.7) -- (0.2, 2.7);
\draw[line width = 1 pt] (-0.1, 2.2) -- (0.2, 2.2);
\draw[line width = 1 pt] (-0.1, 0.8) -- (0.2, 0.8);
\filldraw[cyan!60!white] (0.6,3.2) circle (0.2);
\filldraw[cyan!60!white] (0.2, 3) rectangle (0.6, 3.4);
\node at (0.5,3.2) {\scriptsize $N_{00}$};
\filldraw[cyan!60!white] (0.6,2.7) circle (0.2);
\filldraw[cyan!60!white] (0.2, 2.5) rectangle (0.6, 2.9);
\node at (0.5,2.7) {\scriptsize $N_{01}$};
\filldraw[cyan!60!white] (0.6,2.2) circle (0.2);
\filldraw[cyan!60!white] (0.2, 2.0) rectangle (0.6, 2.4);
\node at (0.5,2.2) {\scriptsize $N_{10}$};
\node at (0.5,1.6) {$\vdots$};
\filldraw[cyan!60!white] (0.6,0.8) circle (0.2);
\filldraw[cyan!60!white] (0.2, 0.6) rectangle (0.6, 1);
\node at (0.5,0.8) {\scriptsize $N_{QQ}$};
\draw[line width = 0.7 pt, style={decorate, decoration=snake}] (0.8,3.2) -- node[above] {\scriptsize $+N^{D}_{00}$} (1.5, 3.2);
\draw[line width = 0.7 pt, style={decorate, decoration=snake}] (0.8,2.7) -- node[above] {\scriptsize $+N^{D}_{01}$} (1.5, 2.7);
\draw[line width = 0.7 pt, style={decorate, decoration=snake}] (0.8,2.2) -- node[above] {\scriptsize $+N^{D}_{10}$} (1.5, 2.2);
\draw[line width = 0.7 pt, style={decorate, decoration=snake}] (0.8,0.8) -- node[above] {\scriptsize $+N^{D}_{QQ}$} (1.5, 0.8);

%% parameter estimation

\draw (1.7, 3.2) -- node[above]{\scriptsize $m_{00}$} (2.4, 3.2);
\draw (1.7,3.2) -- (1.5,3.2);
\draw (1.7, 2.7) --  node[above]{\scriptsize $m_{01}$}(2.4, 2.7);
\draw (1.7,2.7) -- (1.5,2.7);
\draw (1.7, 2.2) -- node[above]{\scriptsize $m_{10}$} (2.4, 2.2);
\draw (1.7,2.2) -- (1.5,2.2);
\node at (2.,1.6) {$\vdots$};
\draw (1.7, 0.8) --  node[above]{\scriptsize $m_{QQ}$} (2.4, 0.8);
\draw (1.7,0.8) -- (1.5,0.8);
\draw (2.4, 3.2) -- (2.4, 0.8);
\draw (2.4,2) -- (2.6,2);

\node[red!70!black] at (2.82,2.01) {$\bar{x}_\mu$};
\draw (2.8,2) circle (0.21);
\end{scope}

\draw (2.7, 2) -- (3.,2);
\begin{axis}[
	at ={(3 cm, 0.5 cm)},
  	domain=0:1.5,
  	samples=50,
  	smooth,
  	width=3 cm,
  	height=4.5 cm,
	axis y line=center,
	axis x line=center,
	ymin=0,
	ymax = 2.35,
	xmin = 0,
	xmax = 1.5,
	restrict y to domain=0:2.35,
	restrict x to domain=0:1.5,
	xtick=\empty,
	ytick=\empty,
	ylabel = $\langle \hat{X} \rangle$,
	xlabel = $d$,
	x axis line style={name path=xaxis},
	y axis line style={name path=yaxis}
]
  \addplot [name path = plot, black] {(1.0685*x^2 + 1.86658* x^4 + 0.633563*x^6 + 0.120831*x^8)*exp(-x^2)};
\addplot[red!30] fill between[of=yaxis and plot,reverse=false, soft clip={domain y = 1.1:1.3}];
\addplot[red!30] fill between[of=xaxis and plot, soft clip={domain = 0.89:0.99}];
\end{axis}

\fill[red!70!black] (3,2) circle (0.05);
\draw[{Bar}-{Bar},red!70!black,thick] (3,1.85) -- (3,2.15);
\node[red!70!black] at (3.5,2.5) {$\Delta X$};
\draw[red!70!black]  (3.3, 2.35) -- (3, 2.15);
\draw[red!70!black,->,thick] (3,2) -- (3.45,2);
\draw[red!70!black,thick] (3.45,2) -- (3.89,2);
\draw[red!70!black,->,thick] (3.89,2) -- (3.89,1.25);
\draw[red!70!black,thick] (3.89,1.25) -- (3.89,0.5);
\fill[red!70!black,thick] (3.89,0.5) circle (0.05);
\node[red!70!black] at (3.89,0.2) {$\tilde{d} \pm \Delta d$};

\draw[blue!70!black,thick] (-4,0) rectangle (1.26,3.7);
\node[blue!70!black] at (-1.375,3.9) {Data acquisition};

\draw[red!70!black,thick] (1.3,0) rectangle (4.5,3.7);
\node[red!70!black] at (2.9,3.9) {Parameter estimation};

\end{tikzpicture}
%\end{document}
\caption{Schematic representation of the estimation procedure. In the data acquisition block, the image of the two sources enters into an (eventually misaligned) demultiplexing device that performs a mode decomposition affected by crosstalk. The intensity of each mode is then measured with noisy detectors. 
Parameter estimation is performed (in post-processing) linearly combining, the measured intensities with optimal coefficients, and comparing the result with a calibration curve.}
\label{Fig:scheme}
\end{figure}
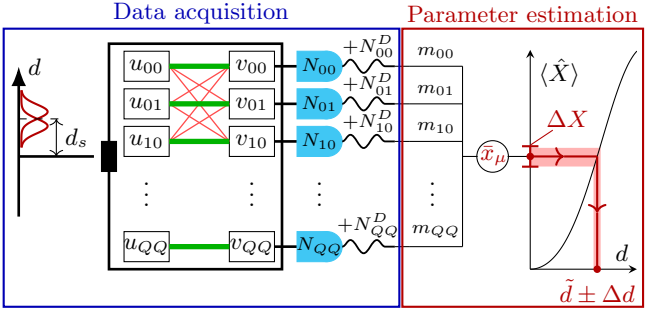

{\it Construction of optimal observable and estimator \:--}
To estimate the distance $d$ between the two thermal sources, we use the method of moments \cite{Luca_Augusto_review}. 
Following this approach an estimator of $d$ is obtained from the sample mean $\overline{x}_\mu = \sum_{i=1}^\mu x_i/\mu$ of a series of $\mu$ independent measurement results $x_i$ of a given observable $\hat{X}$.
In particular, the separation estimator is given by the parameter value $\tilde{d}$ for which the sample mean $\overline{x}_\mu$ equals the expectation value $\langle \hat{X} \rangle$ \footnote{All expectation values are intended with respect to the state of the sources in the image plane $\hat{\rho}(\theta, d)$, e.g. \unexpanded{$ \langle \hat{X} \rangle = \Tr [ \hat{X}  \hat{\rho}(\theta, d)]$}.} of the measurement operator $\hat{X}$.
The dependence of $\langle \hat{X} \rangle$ on the separation $d$ could be known either from theory or from a previous calibration experiment (see the parameter estimation block of Fig.~\ref{Fig:scheme}).
For sufficiently large values of $\mu$, according to the central limit theorem, $\overline{x}_\mu$ is normally distributed with mean value $\langle \hat{X} \rangle$ and variance $( \Delta \hat{X})^2/\mu$. 
Consequently, the estimation error is given by the error propagation formula   $(\Delta d)^2 = ( \Delta \hat{X})^2/\mu \left(\frac{\partial\langle\hat{X}\rangle}{\partial d} \right)^2$.

Let us now assume that we can measure the intensity of $K$ spatial modes $v_k({\bf r})$ with associated field operators $\hat{a}_k$. 
In other words, we have access to the measurement operators $\hat{\bf N} = ( \hat{N}_1, \dots, \hat{N}_K)$, with $\hat{N}_k = \hat{a}^\dagger_k \hat{a}_k$, and we can measure arbitrary linear combinations $\hat{X}_{\tilde{\bf m}} = \tilde{\bf m}\cdot \hat{\bf N}$, with $\tilde{\bf m}$ the measurement-coefficients vector. 
In this case, it is possible to perform an analytical optimization over all possible linear combinations of the accessible operators, i.e. to calculate the measurement sensitivity $M[d,\theta, \hat{\bf N}]  = \max_{\tilde{\bf m}}  \left(\frac{\partial\langle\hat{X}_{\tilde{\bf m}}\rangle}{\partial d} \right)^2/( \Delta \hat{X}_{\tilde{\bf m}})^2$. 
This optimization yields \cite{GessnerPRL2019}
\begin{equation}
M[d,\theta, \hat{\bf N}] = {\bf D}[d,\theta,\hat{\bf N}]^T \Gamma[d,\theta,\hat{\bf N}]^{-1}{\bf D}[d,\theta,\hat{\bf N}],
\label{M}
\end{equation}
with $\Gamma[d,\theta,\hat{\bf N}]$ the covariance matrix with elements $\Gamma_{k,l}[d,\theta,\hat{\bf N}] 
%= {\rm cov}(\hat{N}_k, \hat{N}_l)
=\langle\hat{N}_k \hat{N}_l \rangle -  N_k N_l$ ,
and ${\bf D}[d,\theta,\hat{\bf N}]$ the derivatives vector, with components ${\bf D}_k [d,\theta,\hat{\bf N}] = \frac{\partial N_k}{\partial d}$, where we denoted the mean photon number in the measurement modes as $N_k = \langle \hat{N}_k \rangle$.
The optimum given by Eq. \eqref{M} is obtained for $\tilde{\bf m} = {\bf m}$ \cite{GessnerPRL2019}, with 
\begin{equation}
{\bf m} = \eta \Gamma^{-1}[d, \theta, \hat{\bf N}] D[d, \theta,\hat{\bf N}],
\label{m}
\end{equation}
and $\eta$ a normalization constant. 
The measurement sensitivity $M[d,\theta,\hat{\bf N}]$ obeys the following chain of inequalities $M[d, \theta,\hat{\bf N}] \leq \mathcal{F}[d, \theta, \hat{X}_{\bf m}] \leq \mathcal{F}_Q\left[d, \theta\right]$.
Here, $\mathcal{F}[d, \theta, \hat{X}]$ denotes the Fisher information that bounds the sensitivity of the estimation of $d$ from measurements of $\hat{X}$ according to the Cram\'er-Rao lower bound $(\Delta d)^2  \geq (\mu \mathcal{F}[d, \theta, \hat{X}])^{-1}$ \cite{helstrom1976, BraunsteinCaves1994, holevo2011probabilistic,Paris2009,GiovannettiLoydMaccone,Luca_Augusto_review}.
Finally,  $\mathcal{F}_Q\left[d, \theta\right] = \max_{\hat{X}}  \mathcal{F}[d, \theta, \hat{X}]$ is the quantum Fisher information which determines the ultimate metrological sensitivity \cite{ BraunsteinCaves1994}.

To calculate the quantities \eqref{M} and \eqref{m}, we extend the modes $u_\pm({\bf r})$ to a complete orthonormal basis, and we expand the field operators $\hat{a}_k$ in this basis.
By means of this expansion, the mean photon number $N_k$ and the correlations $\langle \hat{N}_k \hat{N}_l \rangle$ can be fully expressed in terms of the expectation values $\langle \hat{b}^\dagger_\pm \hat{b}_\pm \rangle = N_\pm$, $\langle \hat{b}^\dagger_\pm \hat{b}_\pm \hat{b}^\dagger_\pm \hat{b}_\pm \rangle = 2 N_\pm^2$ and $\langle \hat{b}^\dagger_\pm \hat{b}_\pm \hat{b}^\dagger_\mp \hat{b}_\mp \rangle = N_+ N_-$ \cite{sorelli2021moment}. Finally, we obtain
%\begin{widetext}
\begin{align}
\left(\Gamma[d,\theta,\hat{\bf N}]\right)_{k,l} &= (N\kappa)^2 (|f_{+,k}|^2 |f_{+,l}|^2 +|f_{-,k}|^2 |f_{-,l}|^2  \nonumber \\ &+2\Re \left[ f_{+,k} f_{+,l}^* f_{-,k} f_{-,l}^*\right]) + \delta_{k,l} N_k, \label{Gamma}\\
(D[d,\theta, \hat{\bf N}])_k &= 2 N\kappa \Re\left( f_{+,k}^*\frac{\partial f_{+,k}}{\partial d} + f_{-,k}^*\frac{\partial f_{-,k}}{\partial d}\right), \label{D}
\end{align}
%\end{widetext}
with the mean photon number $N_k =  N\kappa( |f_{+,k}|^2 +|f_{-,k}|^2 )$, and $f_{\pm,k} =  \int d^2 {\bf r} v^*_k ({\bf r}) u_0 ({\bf r} \mp {\bf r}_0)$. 

{\it Optimality of the estimator\:--}
We now study the optimality of this strategy by comparing the measurement sensitivity $M[d,\theta, \hat{\bf N}]$~\eqref{M} with the Fisher information.
To this goal, and for the rest of this work, we consider a Gaussian PSF, $u_0({\bf r}) = \sqrt{2/(\pi w^2)}\exp( - |{\bf r}|^2/w^2)$.
Moreover, let us assume ideal intensity measurements in the Hermite Gaussian (HG) basis $v_k({\bf r}) =  u_{k=(n,m)}({\bf r})$ with $u_{00}({\bf r}) = u_0({\bf r})$.
In this scenario, the covariance matrix \eqref{Gamma} can be inverted analytically and it is possible to derive exact expressions for the measurement sensitivity $M[d,\theta, \hat{\bf N}]$ and coefficients ${\bf m}$ for an arbitrary number $K$ of measured modes \cite{sorelli2021moment}. 
In particular, if we measure intensity in the full HG basis, our estimation strategy saturates the quantum Cram\'er-Rao bound calculated in \cite{LupoPirandola,Nair_2016}, i.e. $ \lim_{K \to \infty} M[d,\theta, \hat{\bf N}] = \mathcal{F}_Q[d,\theta]$ \cite{sorelli2021moment}.
When the number of received photons is low ($N\kappa \ll 1$), the quantum Fisher information is constant $\mathcal{F}_Q[d,\theta] = 2 N \kappa/w^2$ \cite{Tsang_PRX}, accordingly small distances can be resolved as well as large ones.
Moreover, for $N\kappa \ll 1$, the covariance matrix \eqref{Gamma} is essentially diagonal, and we have 
\begin{equation}
M[d,\theta, \hat{\bf N}] \approx \sum_{k=1} \frac{1}{N_k} \left(\frac{\partial N_k}{\partial d} \right)^2,
\label{M_small}
\end{equation}
that coincides with the Fisher information for demultiplexing \cite{Tsang_PRX}, i.e. in the low brightness regime our estimator is efficient for arbitrary $K$.
On the other hand, increasing the sources brightness such that $N\kappa \gtrsim 1$ induces a finite probability of detecting multiple photons in the same mode. 
This fact reduces the quantum Fisher information for intermediate values of $d$, a reduction which is more pronounced the larger $N\kappa$ \cite{LupoPirandola,Nair_2016}.
For thermal sources of arbitrary brightness, an expression for the Fisher information for finite $K$ is unknown.

Let us finally comment that our estimation strategy only requires to access the mean value of a single measurement observable. 
Accordingly, it is practically far more convenient than standard efficient estimators (e.g. maximum likelihood) that require knowledge of the full measurement probability distribution \cite{Helstrom73}. 

{\it Noise sources\:--} 
In the following, we describe how experimental imperfections (whose impact on  intensity measurements is illustrated in the data acquisition block of Fig. \ref{Fig:scheme}) can be included in our model.
\begin{figure}
\includegraphics[width=\columnwidth]{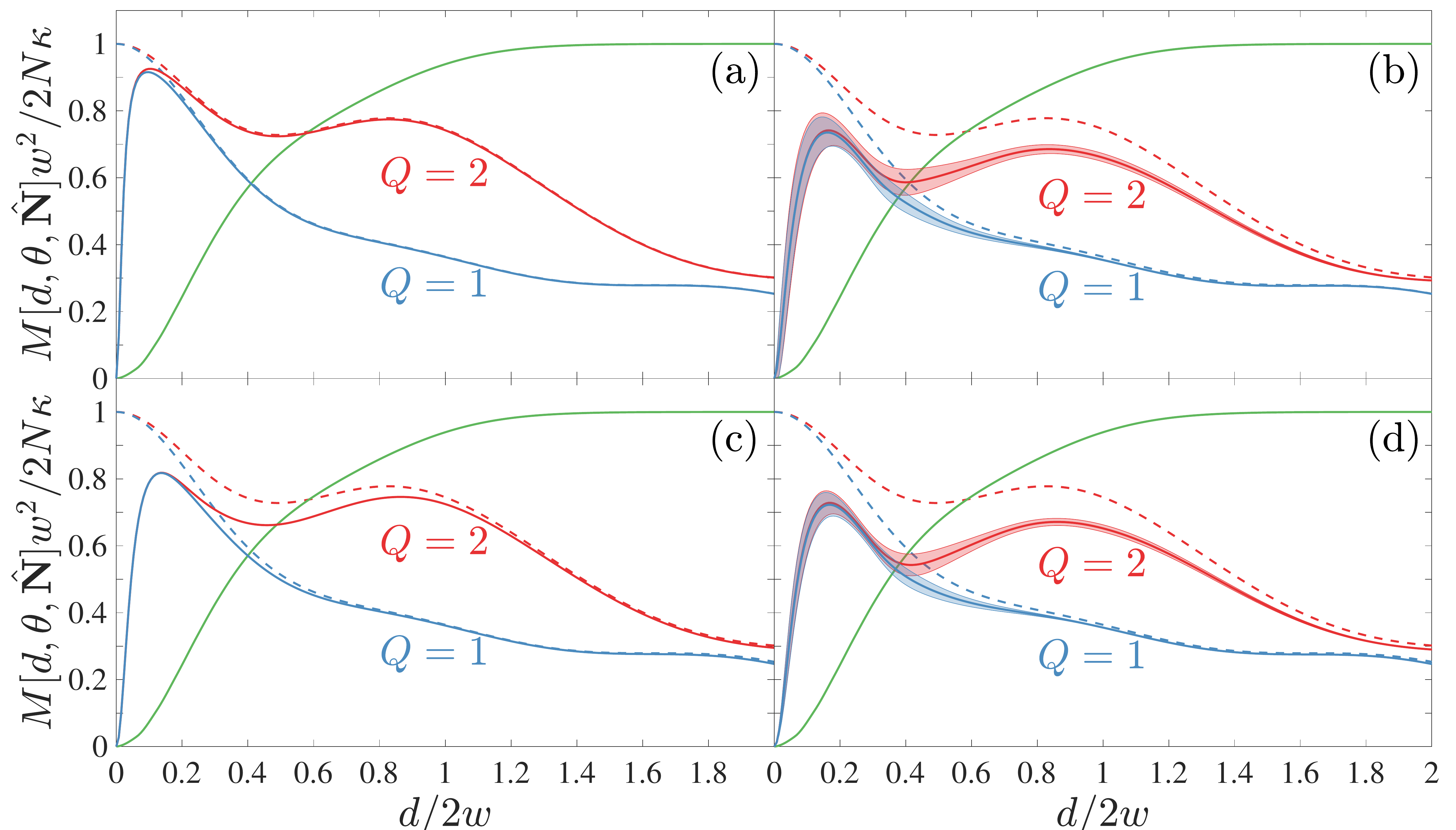}
\caption{Measurement sensitivity $M[d,\theta,\hat{\bf N}]$ for intensity measurements into HG modes $u_{nm}({\bf r})$ with $n,m \leq Q= 1, 2$ ($K = (Q+1)^2$) including (a) misalignment ($d_s/2w = 0.01, \theta_s = \pi/4$), (b) crosstalk ($\langle \overline{|c_{ij}|^2} \rangle = 0.0017$), (c) dark counts ($\sigma_k = 0.001\; \forall k$), and (d) all three imperfections combined.
In panel (b) and (d) solid lines and bands represent the mean and one standard deviation computed over $500$ crosstalk matrices. Dashed lines show the results for ideal measurements, 
while the green solid line describes \added{ideal} direct imaging results \cite{sorelli2021moment}.
For all plots, we assumed $N\kappa = 1.5$ and $\theta = \pi/4$.}
\label{Fig:M}
\end{figure}

First, the assumption that the demultiplexing mode basis $v_k({\bf r})$ is perfectly centred with respect to the centroid of the two sources is often not true in practice.
In particular, an imperfect knowledge of the centroid position leads to a misalignment of the sources and the measurement basis.
A two-dimensional shift ${\bf r}_s =(d_s \cos \theta_s, d_s \sin \theta_s) $ of the sources with respect to the ideal HG measurement basis can be readily included in our model by substituting $f_{\pm,k} =  \beta_k(\pm {\bf r}_0 - {\bf r}_s) =  \int d^2 {\bf r} u^*_k ({\bf r}) u_0 ({\bf r} \mp {\bf r}_0 + {\bf r}_s)$ in Eqs. \eqref{Gamma} and \eqref{D}.
Figure \ref{Fig:M} (a) shows that \replaced{misalignment is mostly relevant when it is of the order of the source separation. On the other hand, for alignment precisions an order of magnitude smaller than the separation,}{alignment must be done with a precision of the order of the source separation. When this is the case,} the impact of misalignment on the measurement sensitivity $M[d,\theta,\hat{\bf N}]$ can be ignored.

Let us now consider the impact of imperfections in the mode decomposition. 
In particular, we model crosstalk between the detection modes as a unitary matrix $c_{kl}$ that maps the ideal HG mode basis $u_l({\bf r})$ into the actual measurement basis $v_k({\bf r}) = \sum_l c_{kl} u_l({\bf r})$. 
In practically relevant scenarios, crosstalk is generally weak, namely the off diagonal elements of the coupling matrix are much smaller than the diagonal ones \cite{Boucher:20, gessner2020}. 
The overlap functions to be used in Eqs. \eqref{Gamma} and \eqref{D} are now given by $f_{\pm,k} = \sum_l c_{kl} \beta_l(\pm {\bf r}_0- {\bf r}_s)$.
To assess the impact of weak crosstalk on our estimator, following \cite{gessner2020}, we numerically generate random $K\times K$ crosstalk matrices $c_{ij}$ resulting into an average crosstalk probability $\langle \overline{|c_{ij}|^2} \rangle = \langle \sum_{i \neq j =1}^K |c_{ij}|^2/K(K-1) \rangle$, where $\langle \cdot \rangle$ represents here an ensemble average \footnote{Here, we assumed $K-$dimensional cross-talk matrices, however by considering matrices of size $D>K$ this model allows to account for losses into modes that are not measured \cite{gessner2020}.}.
Figure \ref{Fig:M} (b) shows the measurement sensitivity $M[d,\theta,\hat{\bf N}]$ averaged over $500$ random crosstalk matrices.

The last noise source we consider is electronic noise at the detection stage, i.e. dark counts.
We model this effect, by adding to the quantum mechanical photon number operators in the measurement modes $\hat{N}_k$ a classical random variable $\xi_k$ which is thermally distributed with mean value $N_k^{\rm dc}$. 
The ratio between the dark counts and the number of received photons $\sigma_k = N_k^{\rm dc}/2N\kappa$ gives the strength of detection noise. 
The mean photon number $N_k^\prime$ in the detection modes as well as the covariance matrix $\Gamma^\prime[d,\theta,\hat{\bf N}]$ are now calculated not only taking quantum mechanical expectation values, but also classical averages over the probability distribution of $\xi_k$.
We therefore obtain $N_k^\prime = N_k + N_k^{\rm dc}$, and since $N_k^{\rm dc}$ does not depend on $d$, the derivative vector \eqref{D} is unaffected by dark counts.
On the other hand, the covariance matrix \eqref{Gamma} acquires an additional diagonal term $\Gamma^\prime[d,\theta,\hat{\bf N}] = \Gamma[d,\theta,\hat{\bf N}] + \delta_{kl} N_k^{\rm dc}(2 N_k^{\rm dc} + 1)$.
The influence of dark counts on the sensitivity $M[d,\theta,\hat{\bf N}]$ is shown in Fig.~\ref{Fig:M} (c), assuming $\sigma_k$ to be the same for all modes.

The combined effect of misalignment, crosstalk and dark counts on the measurement sensitivity $M[d,\theta,\hat{\bf N}]$ is shown in Fig.~\ref{Fig:M} (d) for experimentally relevant imperfections \footnote{We assumed an alignment of the demultiplexer up to $2 \%$ of the PSF diameter ($d_s/2w = 0.01$), the mean crosstalk intensity experimentally measured in \cite{Boucher:20} \unexpanded{$\langle \overline{|c_{ij}|^2} \rangle = 0.0017$}, and an electronic noise  \unexpanded{$ \sigma_k = 0.001$}, which, for  \unexpanded{$N \kappa = 1.5$} is compatible with low dark photon counters.}.
Considering all noise sources together, \replaced{misalignment and crosstalk modify the demultiplexing basis, accordingly they affect the overlap functions $f_{k,\pm}$. On the other hand, dark counts enter at the detection stage and affect only the diagonal of the covariance matrix.}{we have that misalignment and crosstalk affect the overlap functions $f_{k,\pm}$, while dark counts affect only the diagonal of the covariance matrix.}
Accordingly, Eq.~\eqref{M_small}, which coincides with the Fisher information in the $N\kappa \ll 1$ regime, remains valid if we replace $N_k$ with $N_k^\prime$.
As a consequence, in the low brightness regime, our estimator remains efficient even in the presence of noise.

\begin{figure}[t]
\centering
\includegraphics[width=\columnwidth]{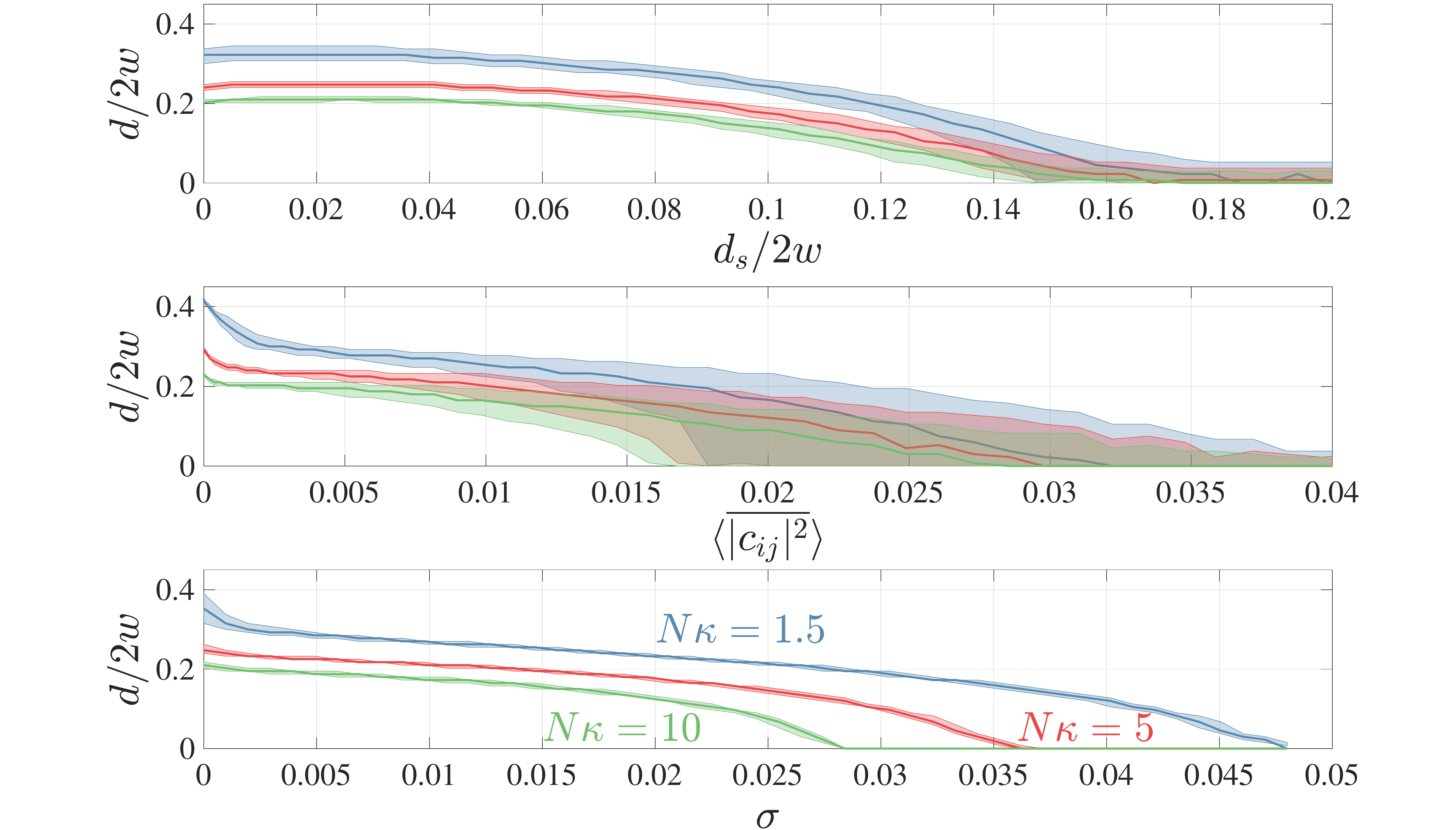}
\caption{Smallest source separation $d$ at which ideal direct imaging outperforms demultiplexing (into HG modes $u_{nm}({\bf r})$ with $n,m \leq 2$) as a function of the (a) misalignment, (b) crosstalk and (c) dark counts strengths for a fixed finite value of the other two imperfections (same as in Fig. \ref{Fig:M}), and different brightnesses $N\kappa = 1.5$ (blue), $5$ (red), $10$ (green).}
\label{Fig:diagram}
\end{figure}

Independently of the number of received photons $N\kappa$, Fig.~\ref{Fig:M} shows that all noise sources cause $M[d,\theta, \hat{\bf N}]$ to vanish for $d \to 0$, and therefore make it harder to resolve small distances.
\replaced{However, even when all imperfections are combined, for small separations, demultiplexing outperforms the most common spatially resolved intensity measurements (direct imaging), which we considered without imperfections.}{However, even when all imperfections are combined, demultiplexing allows to outperform direct imaging for small separations.}
The minimal distance at which ideal direct imaging outperforms imperfect demultiplexing into HG modes $u_{nm}({\bf r})$ with $n,m \leq 2$ (e.g. the crossing point between the red and green curves in Fig. \ref{Fig:M}) is illustrated in Fig.~\ref{Fig:diagram} \footnote{The non monotonous behaviour of $M[d,\theta, \hat{\bf N}]$ for bright sources ($N\kappa \gtrsim 10$) can lead to multiple intersections with the direct imaging curve. Therefore, there are noise regimes where it can be convenient to use demultiplexing to estimate separations larger than those presented in Fig. \ref{Fig:diagram}. Additional figures showing this behaviour are included in \cite{sorelli2021moment}.}. 
Increasing the source brightness allows to resolve smaller distances with both methods, and at the same time reduces the minimal distance at which direct imaging surpass demultiplexing.

\begin{figure}[t]
\includegraphics[width=\columnwidth]{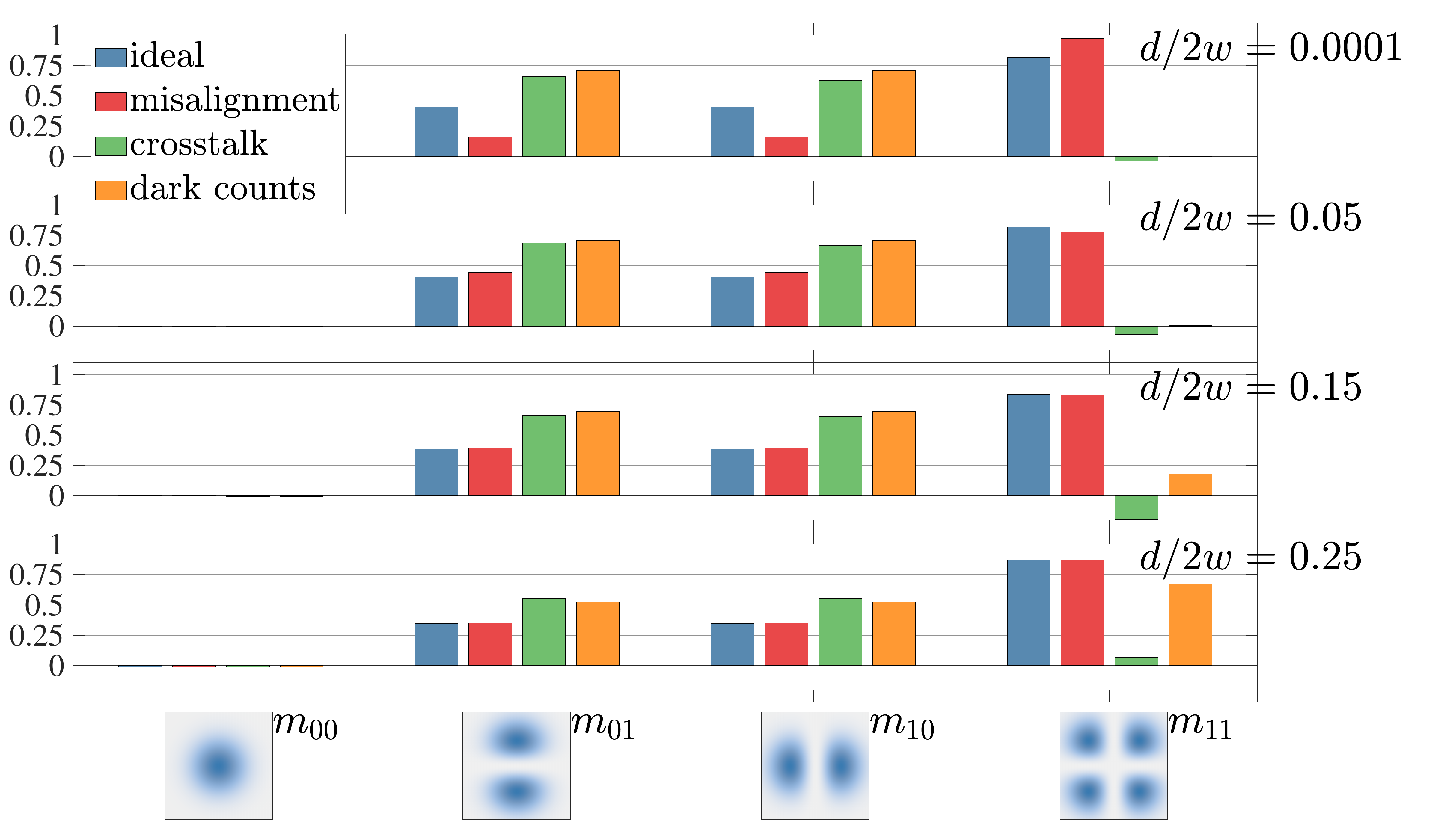}
\caption{Optimal oefficients $m_{ij}$ for measurements in the HG basis $u_{ij}({\bf r})$ with $i, j \leq 1$. The modes' intensity distributions are plotted below the corresponding coefficients. Different noise sources are considered: (blue) none, (red) misalignment ($d_s/2w = 0.01$, $\theta_s = \pi/4$), (green) crosstalk ($\langle \overline{|c_{ij}|^2} \rangle = 0.0017$), (orange) dark counts ($\sigma_k = 0.001\; \forall k$). Green bars for crosstalk are averaged over $500$ crosstalk matrices.  All plots correspond to $N\kappa = 1.5$ and $\theta = \pi/4$.} 
\label{Fig:coefficients}
\end{figure}

{\it Optimal observable\:--} 
Let us finally discuss the observable that practically achieve the sensitivity bounds discussed above.
To this goal, in Fig.~\ref{Fig:coefficients}, we present the coefficients $m_{ij}$, Eq.~\eqref{m}, of the optimal linear combination of intensity measurements in the HG modes $u_{ij}({\bf r})$ with  $i,j \leq 2$ for different source separations $d$.
Comparing the top three panels in Fig.~\ref{Fig:coefficients}, we see that, for small separations and all noise sources, the coefficients are weakly dependent on $d$.
Accordingly, in the, arguably, most interesting range of small separations $d$, the observable that makes the best use of the available measurements does not change with the real value of the parameter.
Further illustrations of this behaviour are provided in \cite{sorelli2021moment}.

Let us now have a look at the amplitudes of the various coefficients.
First, the fundamental mode $u_{00}$ contains no information on $d$ for small separations, accordingly  $m_{00} = 0$.
In the ideal case (blue bars in Fig.~\ref{Fig:coefficients}), for every $k \leq Q$ all coefficients $m_{i,k-i}$ (with $i \leq k$) are degenerate, and their amplitude increase with $k$.
In fact, in the absence of noise, the optimal observable amplifies the small signals in the higher order modes to extract the most information on the parameter out of them.
Different noise sources modify this behaviour. 
Misalignment (red bars in Fig.~\ref{Fig:coefficients}) is influential for $d \lesssim d_s$ and tends to increase higher-order coefficients.
On the contrary, both crosstalk and dark counts (green and orange bars in Fig.~\ref{Fig:coefficients}) add noise to the higher order modes.
Accordingly, the coefficients for these modes get strongly depleted, and for small separation the ultimate sensitivity can be achieved by only measuring $u_{01}({\bf r})$ and $u_{10}({\bf r})$.
For larger separations, even in presence of noise, the optimal observable gets significant contributions also from higher order modes (e.g. orange bars in the bottom panel of Fig.~\ref{Fig:coefficients}).

{\it Conclusions\:--} 
We presented a procedure to estimate the separation between two thermal sources from an optimal combination of demultiplexed intensity measurements.
In the limiting case of ideal intensity measurements into infinitely many HG modes, our approach reaches the quantum Cram\'er-Rao bound for arbitrary source brightness and separations.
For faint sources, the sensitivity $M[d,\theta,\hat{\bf N}]$ of our method saturates the Fisher information even in the noisy scenario.
In other words, in the $N\kappa \ll 1$ regime, our estimator makes the best possible use of a (possibly noisy) demultiplexing measurement.
Interestingly, this approach allows to reach optimality using a practical estimator that only requires knowing the mean value of a single observable.
In addition, for small separations $d$, the weak dependence of the optimal observable on $d$ allows to identify an estimation strategy valid over a wide range of separations.

The optimal observable is given by Eq. (4) and can be determined in practice in different ways. A measurement of the covariance matrix and derivative vector using test sources naturally accounts for all experimental noise sources. Alternatively, these properties can be predicted from theory models if sufficient information about the noise is known: 
%from an independent characterization of the setup.
Detection noise is routinely measured in experiments and the crosstalk matrix can be extracted from a careful characterization of the mode sorting device. Misalignment errors mostly stem from an imprecise knowledge of the source centroid and can be contained by scanning the demultiplexer in the image plane or using adaptive strategies \cite{Grace:20}.

Finally, we point out that %with a fixed apparatus
different measurement coefficients can be chosen at the estimation stage after the measurements have been performed. Therefore, even in presence of a dynamically changing parameter, our approach allows to select at any time the optimal observable in post processing.

{\it Acknowledgement\:--} GS acknowledges financial support of ONERA - the French aerospace lab.  MG acknowledges funding by the LabEx ENS-ICFP:ANR-10-LABX-0010/ANR-10-IDEX-0001-02 PSL*. This work was partially funded by French ANR under COSMIC project (ANR-19-ASTR-0020-01). This work received funding from the European Union’s Horizon 2020 research and innovation programme under grant agreement No 899587. This work was supported by the European Union’s Horizon 2020 research and innovation programme under the QuantERA programme through the project ApresSF.
\bibliography{SR}{}

\end{document}